\documentclass[journal]{IEEEtran}
 
\IEEEoverridecommandlockouts
\usepackage{amsmath,amssymb,amsthm}
\usepackage{algorithm}
\usepackage{algorithmic}
\usepackage{booktabs}
\usepackage[margin=0.75in,columnsep=0.25in]{geometry}
\usepackage{times}
\usepackage{hyperref}
\usepackage{bm}
\usepackage{color}
\usepackage{graphicx}
\usepackage{cite}

\newcommand{\bSigma}{\boldsymbol{\Sigma}}

\newcommand{\bPsi}{\boldsymbol{\Psi}}

\newcommand{\C}{\mathbb{C}}

\newcommand{\vect}[1]{\bm{#1}}
\newcommand{\cG}{\mathcal{G}}
\newcommand{\cK}{\mathcal{K}}
\newcommand{\cS}{\mathcal{S}}
\newcommand{\cT}{\mathcal{T}}

\newcommand{\unvec}{\operatorname{unvec}}

\begin{document}
 
\title{2D Block OMP for Near-Field Channel Estimation with Extremely Large-Scale UPA
\thanks{This work was supported by the Natural Science Foundation of Sichuan Province under Grant 2026NSFSC0401.}
}

\author{
\IEEEauthorblockN{
 Hongwei Wang$^{\star}$, Shuai Wang$^{\star}$, Xiaofei Long$^{\star}$, Yongjun Chen$^{\star}$, and Mangang Xie$^{\dagger}$
}
\\
\IEEEauthorblockA{
\textit{$^{\star}$University of Electronic Science and Technology of China, Chengdu, China} \\
\textit{$^{\dagger}$Northwest Normal University, Lanzhou, China}\\
\{hongwei\_wang,\ shuaiwang\}@uestc.edu.cn; xiemg@nwnu.edu.cn \\
}
}

\maketitle

\begin{abstract}
Channel estimation for extremely large-scale uniform planar arrays (UPAs) requires exploiting near-field channel structure to reduce pilot overhead. With a specifically devised channel representation matrix, the near-field channel exhibits a 2D sparse structure, which a conventional 1D block partition might divide into multiple groups. This work develops a 2D block orthogonal matching pursuit (2D-BOMP) estimator that exploits this structure through rectangular coefficient grouping. We establish sufficient recovery conditions for noiseless and noisy measurements and compare the sufficient pilot-length requirements of 1D and 2D partitions. Numerical simulations demonstrate the estimation benefits of 2D-BOMP in estimating the near-field channels of large-scale UPAs, compared with the conventional 1D-BOMP as well as the polar-domain based OMP method.
\end{abstract}


\section{Introduction}

Extremely large-scale antenna arrays (ELAAs) enable high beamforming gains to compensate for severe propagation losses in millimeter-wave and terahertz communications~\cite{cui2022near,ye2024extremely,wang2024fast}. As the array aperture increases, the near-field region can extend to ten to hundred meters~\cite{ding2025near}, leading the users (UE) may be located into the near-field region of the base station (BS). In this region, spherical wavefronts make the channel depend jointly on propagation angles and distance, rendering conventional angle-only far-field models inadequate and challenging the channel estimation.

Exploiting near-field channel sparsity offers a promising solution. A polar-domain based near-field channel estimation approach~\cite{cui2022channel} (and its subsequent works) is therefore presented, in which the near-field channel is sparsely represented in the polar domain, and orthogonal matching pursuit (OMP) can then be utilized to estimate it. However, the polar-domain method suffers from excessively large dictionary atoms that are non-orthogonal, which can severely degrade the performance of compressive sensing recovery. To deal with this challenge, in our previous work~\cite{wang2023compressive,wang2025near}, we proposed a modified DFT channel representation approach, in which the near-field channel of a uniform linear array (ULA) exhibits block-sparse structure, enabling structured recovery using block OMP (BOMP) or pattern-coupled sparse Bayesian learning (PCSBL).

Extending this concept to the uniform planar array (UPA) case, which is more general in practice, requires replacing the 1D spatial model with a 2D spatial framework. Consequently, the near-field channel features 2D clustered coefficients across both the azimuth and elevation domains. However, applying a conventional 1D block partition to the vectorized coefficients may divide a single path-associated cluster into several non-contiguous blocks. Although vectorization preserves the coefficients, such a partition fails to explicitly exploit their two-dimensional organization.

In our recent work~\cite{chen2026near}, we characterized the two-dimensional block-sparse structure of UPA near-field channels using modified DFT dictionaries and developed a two-dimensional PCSBL estimator to exploit this structure. The simulation results show that exploiting the two-dimensional block sparse structure indeed improves channel estimation accuracy. However, a theoretical analysis quantifying the performance gains of two-dimensional grouping over a one-dimensional partition remains missing. Therefore, in this work, we first develop a 2D-BOMP-based channel estimator, establish sufficient recovery conditions for the noiseless and noisy cases, and compare the sufficient pilot-length requirements of one-dimensional and two-dimensional grouping. Numerical simulations further illustrate the advantage of employing the 2D block sparse structure when estimating the near-field channel of the UPA.

\section{Problem Formulation}

\subsection{System Model}
We consider a downlink transmission scenario. A BS equips with an ELAA configured as a UPA of $N=N_xN_y$ elements and serves single-antenna users located within the Rayleigh distance of the array. This is practically justified since the ELAA, especially when operating at high-frequency bands like THz, extends the Rayleigh distance up to hundreds of meters. In this region, the far-field planar wavefrount assumption is no longer applicable. Therefore, we adopt the spherical wavefront model, resulting a near-field channel that jointly depends on both the distance and the angle. Specifically, the channel $\bm H\in \mathbb{C}^{N_y\times N_x} $ between the BS and a given user is modeled as
\begin{align}
    \bm H = \sum_{l=0}^{L-1} \tilde g_l\, \bm A(r_l,\theta_l,\phi_l)
\label{eq:H}
\end{align}
where $L$ is number of propagation paths, $\tilde g_l$ is the complex path gain of the $l$th path, and $\bm A(r,\theta,\phi)$ is the near-field UPA steering matrix parameterized by distance $r$, elevation $\theta$, and azimuth $\phi$. Its $(n_y,n_x)$th entry is given by
\begin{align}
    &\left[\bm A(r,\theta,\phi)\right]_{n_y,n_x}
=\notag\\
&\qquad\qquad \frac{1}{\sqrt{N}}
\exp\left[
-\frac{\mathrm{j}2\pi}{\lambda_c}
\left(
r^{(n_y,n_x)}(r,\theta,\phi)-r
\right)
\right]
\label{eq:steering_entry}
\end{align}
for \(n_y=0,\ldots,N_y-1\) and \(n_x=0,\ldots,N_x-1\), where \(\lambda_c\) is the carrier wavelength and \(r^{(n_y,n_x)}(r,\theta,\phi)\) denotes the propagation distance from the corresponding source point to the \((n_y,n_x)\)th array element. Note that $r^{(0,0)}(r,\theta,\phi) \equiv r$.

Since $\bm N$ is typically on the order of thousands of elements in mmWave/THz systems, directly estimating every entry of $\bm H$ like the least square solution would incur prohibitive pilot overhead. 
Therefore, the channel's inherent structure should be exploited to reduce this overhead. As shown in our prior work~\cite{chen2026near}, a first-order Fresnel approximation decomposes \(\bm A(r,\theta,\phi)\) into the outer product of two near-field uniform linear array steering vectors, one per array dimension, each of which admits a block-sparse representation under a modified-DFT dictionary. Combining these two facts yields
\begin{equation}
\bm H = \bm D_y\bm \Sigma\bm D_x^T,
\label{eq:HDSigmaD}
\end{equation}
where $\bm D_x\in\C^{N_x\times N_x}$ and $\bm D_y\in\C^{N_y\times N_y}$ are the modified-DFT dictionaries for the two array dimensions. Explicitly, for the horizontal dimension,
\begin{equation}
\bm D_x = \mathrm{diag}\big(\bar{\bm b}(\zeta_a,r)\big)\,\bm W_x,
\label{eq:Dx}
\end{equation}
where $\bm W_x\in\C^{N_x\times N_x}$ is a unitary DFT matrix, and $\bar{\bm b}(\zeta,r)\in\C^{N_x}$ carries the near-field (Fresnel) quadratic phase correction, with $n$th entry
\begin{equation}
[\bar{\bm b}(\zeta,r)]_n=\frac{1}{\sqrt{N_x}}\exp\!\left(-j\frac{2\pi}{\lambda_c}\frac{n^2d^2}{2r}(1-\zeta^2)\right);
\label{eq:bbar}
\end{equation}
$\mathbf D_y$ is defined analogously. With $\bm D_x$ and $\bm D_y$, $\bm \Sigma$ can be represented as~\cite{chen2026near}
\begin{equation}
\bm \Sigma = \sum_{l=0}^{L-1}\tilde g_l\,\boldsymbol\beta_{y,l}\boldsymbol\beta_{x,l}^T,
\label{eq:Sigma}
\end{equation}
with $\boldsymbol\beta_{y,l}\in\C^{N_y}$ and $\boldsymbol\beta_{x,l}\in\C^{N_x}$ the dictionary-domain coefficient vectors of the $l$th path. As pointed out in~\cite{wang2025near}, $\boldsymbol\beta_{y,l}$ and $\boldsymbol\beta_{x,l}$ are block-sparse vectors. Concretely, partition $\{0,\dots,N_y-1\}$ into $S_y\triangleq N_y/B_y$ contiguous blocks $\{K_{y,1},\dots,K_{y,S_y}\}$ of size $B_y$ each, and $\{0,\dots,N_x-1\}$ analogously into $S_x\triangleq N_x/B_x$ blocks $\{K_{x,1},\dots,K_{x,S_x}\}$ of size $B_x$. Because $\boldsymbol\beta_{y,l}$ is supported entirely within a single such block, there is a block index $q(l)\in\{1,\dots,S_y\}$ with $\mathrm{supp}(\boldsymbol\beta_{y,l})\subseteq K_{y,q(l)}$, and likewise a block index $p(l)\in\{1,\dots,S_x\}$ with $\mathrm{supp}(\boldsymbol\beta_{x,l})\subseteq K_{x,p(l)}$, i.e., each rank-one summand in \eqref{eq:HDSigmaD} occupies the contiguous $B_y\times B_x$ tile $K_{y,q(l)}\times K_{x,p(l)}$ of $\bm \Sigma$. It therefore shows that the full representation matrix satisfies
\begin{equation}
\mathrm{supp}(\bSigma)=\bigcup_{l=0}^{L-1}K_{y,q(l)}\times K_{x,p(l)},
\label{eq:support}
\end{equation}
i.e., $\bSigma$ is 2D block-sparse, with at most $L$ active $B_y\times B_x$ tiles.

\subsection{Problem Formulation}

During channel estimation, the BS's pilot signal is precoded via hybrid beamforming prior to transmission, so that the received signal at the user at the $t$th pilot instant is
\begin{align}
 y(t) = \bm h^T \bm f(t) +n(t)
\end{align}
where $\bm h\triangleq\mathrm{vec}(\bm H)$, $\bm f(t)$ denotes the effective per-instant sensing vector jointly determined by the pilot symbol and the hybrid-beamforming precoder, and $n(t)$ is additive white Gaussian noise. Stacking the $T$ pilot observations gives the matrix-form model
\begin{equation}
\bm y = \bm  F\bm h + \bm n,
\label{eq:yFh}
\end{equation}
with $\bm y \triangleq[y(1),\dots,y(T)]^T$, $\bm  F\triangleq[\bm f(1),\dots,\bm f(T)]^T\in\C^{T\times N}$, and $\bm n\triangleq[n(1),\dots,n(T)]^T$.

Substituting the sparse representation \eqref{eq:HDSigmaD} of $\bm H$ and applying the vectorization identity $\mathrm{vec}(\bm A\bm X\bm B^T)=(\bm B\otimes\bm A)\mathrm{vec}(\bm X)$, the received signal is equivalently expressed in the sparse coefficient domain as
\begin{align}
    \bm y & = \bm F \text{vec}(\bm H) + \bm n\notag\\
    & = \bm F (\bm D_y \otimes \bm D_x) \text{vec}(\bm \Sigma) + \bm n\notag \\
    & = \bm \Psi   \bm x + \bm n
\end{align}
where $\bPsi\triangleq\mathbf F(\mathbf D_x\otimes\mathbf D_y)$ is the effective sensing matrix and $\bm x\triangleq\mathrm{vec}(\bm \Sigma)$ is the 2D block-sparse coefficient vector to be recovered. Given the a priori knowledge that $\bm x$ is 2D block-sparse, channel estimation reduces to a structured sparse-recovery task: recover the $L$-tile support and coefficients of $\bm \Sigma$ from $\bm y$, using as few pilots $T$ as the underlying 2D structure allows. In the following, we derive a channel estimation scheme tailored to this structure, together with its supporting theoretical guarantees.

\section{Proposed Method}
In this section, we discuss how to use the block-OMP to estimate the near-field channel for the UPA. The proposed method changes the grouping of the nonzero element set, not the greedy mechanism within the OMP algorithm. Let $\mathcal{G}=\{G_1,\ldots,G_S\}$ be a partition of the $N$ coefficient indices. Algorithm~\ref{alg:bomp} is standard BOMP over this generic partition~\cite{eldar2010block}. 

\begin{algorithm}[t]
\caption{BOMP over a partition $\cG$}
\label{alg:bomp}
\begin{algorithmic}[1]
\REQUIRE $\vect y,\vect\Phi,\cG$, iteration budget $K$
\STATE $\vect r^{(0)}\leftarrow\vect y$, $\cS^{(0)}\leftarrow\varnothing$
\FOR{$k=1,\ldots,K$}
 \STATE $G^{(k)}\leftarrow
 \arg\max_{G\in\cG\setminus\cS^{(k-1)}}
 \|\vect\Phi_G^H\vect r^{(k-1)}\|_2$
 \STATE $\cS^{(k)}\leftarrow\cS^{(k-1)}\cup\{G^{(k)}\}$
 \STATE $\widehat{\vect x}_{\cS^{(k)}}
 \leftarrow\vect\Phi_{\cS^{(k)}}^{\dagger}\vect y$
 \STATE $\vect r^{(k)}\leftarrow
 \vect y-\vect\Phi_{\cS^{(k)}}\widehat{\vect x}_{\cS^{(k)}}$
\ENDFOR
\RETURN $\widehat{\vect\Sigma}_{\rm B}=\unvec(\widehat{\vect x})$
\end{algorithmic}
\end{algorithm}

One intuitively approach to save the near-field channel estimation for UPAs is, like the frequently utilized in estimating far-field channels, to directly vectorize $\bm \Sigma$, resulting in partitioning $\bm x$ into contiguous chunks of length $B_y$, i.e., $\cG_{\rm 1D}=\{H_m\}_{m=1}^{N/B_y}$ with $H_m\triangleq \{(m-1)B_y+1,\ldots,(m)B_y\}$, and $N/B_y$ is assumed to be an integral without loss of generality. This choice is the natural one if $\bm \Sigma$ is treated as a generic vector rather than as a matrix with two-dimensional structure. Because $\bm x$ is a column-major vectorization, each block $H_m$ isolates the $N_y$ (row) dimension and spans only a single column of $\bSigma$. Consequently, this partition overlooks the sparse connections across columns. To take the sparse connections across columns into consideration, we here utilize the following 2D partition, i.e., $\cG_{\rm 2D}=\{G_{p,q}\}$ where $ G_{p,q}\triangleq \{i+jN_y:i\in\cK_{y,q},\ j\in\cK_{x,p}\}$. Each element of $\cG_{\rm 2D}$ is exactly the vectorization of a
$B_y\times B_x$ tile. 

An active $B_y\times B_x$ tile intersects between one and $B_x$ blocks
of $\mathcal G_{\rm 1D}$. If the $L$ active tiles are disjoint and every
tile contains at least one nonzero coefficient in each of its $B_x$
columns, then $K_{\rm 1D}=LB_x$, whereas $K_{\rm 2D}=L$. Otherwise, the
exact active-block counts are determined by the numbers of partition
members intersecting $\operatorname{supp}(\bm{\Sigma}_{\rm B})$ and
may be smaller because of partial tile occupancy or overlap among the
active tiles. This result follows from column-major vectorization. Consequently, full column occupancy causes each tile to
intersect $B_x$ 1D blocks, and disjointness ensures that these blocks are
distinct across paths; in contrast, the geometry-aligned 2D partition
represents each complete tile as a single block.

After obtaining $\hat{\bm x}$ by the BOMP algorithm, the estimated
two-dimensional coefficient matrix is then constructed via $\hat{\bm{\Sigma}}_{ B}
=\operatorname{unvec}(\hat{\bm x})$. As a result, the near-field
channel is reconstructed by
\begin{align}
\hat{\bm H}
=\tilde{\bm D}_{y}\hat{\bm{\Sigma}}_{ B}
 \tilde{\bm D}_{x}^{T}
\end{align}
For the sake of distinction, we hereinafter use 1D‑BOMP and 2D‑BOMP to denote the BOMP‑based channel estimation algorithms for the two different segmentation cases.

\section{Recovery and Pilot-Length Analysis}
\label{sec:analysis}

In this section, we present a comparative theoretical analysis of the
1D-BOMP- and 2D-BOMP-based channel estimation schemes. We consider the full-occupancy and disjoint-tile case, which constitutes
the worst case for the 1D partition in terms of active-block count and
produces the largest support-count gap between 1D-BOMP and 2D-BOMP. Under this assumption, the
sparsity level, the number of coefficients per block, and the total number
of candidate blocks are summarized in Table~\ref{table1}. Both methods
recover the same number of active coefficients, namely,
$P\triangleq kd=LB_xB_y$. In the following, we first compare their recovery
guarantees in the noiseless and noisy settings and then analyze their
required pilot lengths.

\begin{table}[!t]
\centering
\caption{Key parameters in these two algorithms}
\begin{tabular}{cccc}
\hline
        & $k$ & $d$ & $S$ \\
        \hline
1D-BOMP & $LB_x$  & $B_y$  & $N/B_y$ \\
2D-BOMP & $L$  & $B_xB_y$  & $N/B_y$ \\
\hline
\end{tabular}
\label{table1}
\end{table}

\subsection{Noiseless and Noisy Recovery}

We here depend on the block-RIP condition to compare with the recovery guarantee of these two BOMP based channel estimation methods. The sensing matrix $\bm \Phi$ is said to satisfy block-RIP of order $k$ with constant $\delta_k^B$ if
\begin{equation}
 (1-\delta_k^B)\|\vect x\|_2^2
 \leq\|\vect\Phi\vect x\|_2^2
 \leq(1+\delta_k^B)\|\vect x\|_2^2
 \label{eq:brip}
\end{equation}
for every block $x$ sparse vectors $\bm z$. In noiseless case, the following theorem holds:

\textit{Theorem 1 (noiseless BOMP recovery~\cite{li2018new}).}
If $\vect x$ is block-$k$ sparse and
\begin{equation}
 \delta_{k+1}^{B}<\frac{1}{\sqrt{k+1}},
 \label{eq:ripthreshold}
\end{equation}
then BOMP recovers $\vect x$ exactly in $k$ iterations from noiseless
measurements.

Theorem~1 provides a sufficient, but not necessary, condition for exact
BOMP recovery. According to the different setting of the 1D-BOMP and 2D-BOMP based channel estimation methods, their sufficient condistions are, respectively,
\begin{equation}
 \delta_{LB_x+1}^{B,{\rm 1D}}<\frac{1}{\sqrt{LB_x+1}},\qquad
 \delta_{L+1}^{B,{\rm 2D}}<\frac{1}{\sqrt{L+1}}.
 \label{eq:twothresholds}
\end{equation}
Generally, we have $B_x$ is larger than $1$. Therefore, the right-hand threshold for the 2D model is larger and hence less restrictive. It should be noted that the comparison in~\eqref{eq:twothresholds} concerns the required thresholds, rather the actual constants. Therefore, \eqref{eq:twothresholds} alone neither proves that the
2D constant is smaller nor establishes a deterministic pilot reduction.

In the noisy case,~\eqref{eq:ripthreshold} alone is insufficient for support
recovery. Let $\|\vect e\|_2\leq\epsilon$ and let $\cT$ be the active block
set. BOMP with residual stopping rule $\|\vect r^{(k)}\|_2\leq\epsilon$
recovers $\cT$ in $k$ iterations if~\cite{li2018new}
\begin{align}
 \min_{G\in\cT}\|\vect x_G\|_2
 &>\frac{\epsilon}{\sqrt{1-\delta_{k+1}^{B}}}
 +\frac{\epsilon\sqrt{1+\delta_{k+1}^{B}}}
 {1-\sqrt{k+1}\,\delta_{k+1}^{B}}.
 \label{eq:noisycondition}
\end{align}
For sake of simplicity, define the
required block-amplitude margin
\begin{equation}
 \tau(k,\delta,\epsilon)\triangleq
 \frac{\epsilon}{\sqrt{1-\delta}}+
 \frac{\epsilon\sqrt{1+\delta}}{1-\sqrt{k+1}\,\delta}.
\end{equation}
Then, the sufficient noisy conditions for the two methods become
\begin{align}
 \min_{G\in\cT_{\rm 2D}}\|\vect x_G\|_2
 &>\tau\!\left(L,\delta_{L+1}^{B,{\rm 2D}},\epsilon\right),\\
 \min_{G\in\cT_{\rm 1D}}\|\vect x_G\|_2
 &>\tau\!\left(LB_x,\delta_{LB_x+1}^{B,{\rm 1D}},\epsilon\right).
\end{align}

For fixed $\delta$, the margin increases with both $\epsilon$ and $k$; it
also diverges as $\delta$ approaches $1/\sqrt{k+1}$. Hence the smaller 2D
order relaxes the amplitude requirement when the two relevant RIP constants
are comparable. In addition, the unknown non-zero elements in these two methods satisfy
\begin{align}
    \|\vect x_{G_l}^{\rm 2D}\|_2^2
 =\sum_{j=1}^{B_x}\|\vect x_{l,j}\|_2^2
\end{align}
Apparently, $\|\vect x_{G_l}^{\rm 2D}\|_2^2$ provides an
aggregation gain between $1$ and $\sqrt{B_x}$ relative to the strongestactive 1D block. Moreover, because 1D-BOMP must detect each active 1D block separately, a
weak block may limit complete support recovery, whereas 2D-BOMP can
identify the corresponding 2D block using its aggregated energy.

\subsection{Pilot-Length Scaling under Sub-Gaussian Training}

To characterize the required pilot length, we, as most theoretical analysis works, assume that the rows of $\vect F$ are independent, normalized, isotropic sub-Gaussian random vectors. This general model includes Gaussian and various bounded random training designs. The sensing matrix in our channel estimation problem $\vect\Phi=\vect F(\widetilde{\vect D}_x\otimes
\widetilde{\vect D}_y)$ has independent isotropic sub-Gaussian rows since multiplication by the unitary matrix preserves the
isotropy and sub-Gaussian properties of each row. We begin the pilot length analysis with the theorem from Ref.~\cite{eldar2010block}.

For $S$ candidate blocks of size $d$, a normalized
sub-Gaussian sensing matrix satisfies block-RIP of order $k$ with
probability at least $1-\eta$ if
\begin{equation}
T\gtrsim\delta^{-2}
\left[
kd\log\left(\frac{12}{\delta}\right)
+k\log\left(\frac{2S}{k}\right)
+\log\left(\frac{1}{\eta}\right)
\right] 
\label{eq:samplecomplexity}
\end{equation}
In this equation, the first term accounts for the dimension of each support subspace,
the second for uncertainty in the support locations, and the third
for the required reliability. 

Define $ A_{\delta}\triangleq
P\log\left({12}/{\delta}\right)+\log(1/\eta)$ and $
q\triangleq\log\left( {2N}/{P}\right)$. Given a common RIP target $\delta$, failure probability $\eta$, and normalized sensing ensemble, we have 
\begin{align}
\overline T_{\rm 1D}
&\gtrsim  \delta^{-2}
  \left[A_{\delta}+LB_xq\right] \label{eq:pilotcomparison1}\\
\overline T_{\rm 2D}
&\gtrsim \delta^{-2}
  \left[A_{\delta}+Lq\right].
\label{eq:pilotcomparison2}
\end{align}
The shared term $A_{\delta}$ is identical because both methods have the same active-subspace dimension and reliability requirement. Their difference is therefore located to terms of $L B_x q$ and $Lq$. Since $B_x$ is genearlly greater than $1$, the constructed sufficient bound is smaller for the
2D model. Specifically, joint selection reduces the support-location
term from $LB_xq$ to $Lq$. The relative benefit is greater when
support-location uncertainty dominates the shared dimension and
reliability terms. It should be noted that these bounds characterize a theoretical advantage of the structured model rather than an exact reduction in the
minimum number of pilots. This is because the two
methods require different RIP orders and may use different targets,
the reduction in~\eqref{eq:pilotcomparison1} and~\eqref{eq:pilotcomparison2} should not be interpreted
as a universal pilot-saving guarantee for BOMP. Nevertheless, the less demanded pilot length of the 2D-BOMP based channel estimation is verified in next numerical simulation section.

\subsection{Summary}

The preceding recovery and pilot-length analyses demonstrate that the
advantage of 2D-BOMP originates from its geometry-aligned treatment of
each propagation path. In the transformed channel domain, the nonzero
coefficients associated with one path are concentrated within a
$B_y\times B_x$ tile. The proposed method treats this entire tile as one
2D block, whereas 1D-BOMP divides it into as many as $B_x$ column-wise
blocks and detects these blocks separately. Under the full-occupancy and
disjoint-tile assumptions, this distinction reduces the effective
block-sparsity level from $LB_x$ to $L$. Consequently, 2D-BOMP has a less
restrictive block-RIP threshold and can use the aggregate energy of all
coefficients associated with a path for block detection. These benefits
remain conditional on the actual block-RIP constants of the two
partitions.

The same grouping principle explains the pilot-length difference.
Although both methods recover the same $P=LB_xB_y$ active coefficients,
1D-BOMP treats the $B_x$ blocks associated with one path as independently
located supports. In contrast, 2D-BOMP constrains them to be selected
jointly as one candidate tile, thereby reducing the uncertainty in the
support locations. As a result, the support-complexity term decreases
from $LB_xq$ to $Lq$ in the sufficient pilot-length scaling law. Therefore,
the theoretical advantage of 2D-BOMP does not arise from reducing the
number of unknown channel coefficients, but from exploiting the additional
support information encoded by its block partition. In such a partition, the coefficients generated by each physical path form one structured
two-dimensional group rather than several unrelated one-dimensional
blocks.

\section{Numerical Simulation}

In this section, we compare the proposed 2D-BOMP-based channel estimator with the conventional 1D-BOMP-based estimator. We also consider the OMP based channel estimation method in the polar-domain (polar-domain OMP)~\cite{cui2022channel} as a baseline. All these BOMP based methods operate on the same modified-DFT channel representation. The 1D-BOMP baseline selects consecutive groups of six coefficients from the vectorized representation, while the 2D-BOMP implementation uses rectangular $2\times3$ groups. The channel representation matrix in the polar-domain OMP is generated by the guidance given in~\cite{cui2022channel}.

We consider a $32\times256$ UPA with half-wavelength element spacing operates at $f_c=100$~GHz. We set $L=3$ with one LOS path and two NLOS paths. The angular parameters $\theta_l$ and $\phi_l$ are independently drawn from $\mathcal U[-\pi/3,\pi/3]$, and the path distances are sampled from $\mathcal U[2.73,99.84]$~m. Each NLOS path has an average power $13$~dB below that of the LOS component. The entries of the sensing matrix $\bm F$ are i.i.d drawn from $\mathcal{N}(0,1)$. The noise is circularly symmetric complex Gaussian, and its variance is assumed known to all estimators. The signal-to-noise ratio is defined as $\|\bm F \bm h\|^2_2/\|\bm n\|^2_2$. The modified DFT matrix is constructed with $r_0=10$~m and $\theta_0=\phi_0=\pi/12$. The normalized mean square error (NMSE) is considered as a metric to evaluate the accuracy of the channel estimate, which is defined as
\begin{align}
    \mathrm{NMSE}=\mathbb{E}\left(
    {\|\hat{\bm H}  -\bm H  \|_F^2}/
    {\| \bm H \|_F^2}
    \right)
\end{align}
where $\hat{\bm H}$ and $\bm H$, are, respectively, the estimated and the true value of the channel. The results are evaluated via $1000$ independent Monte Carlo runs.

\begin{figure}[htbp]
  \centering
  \includegraphics[width=0.35\textwidth]{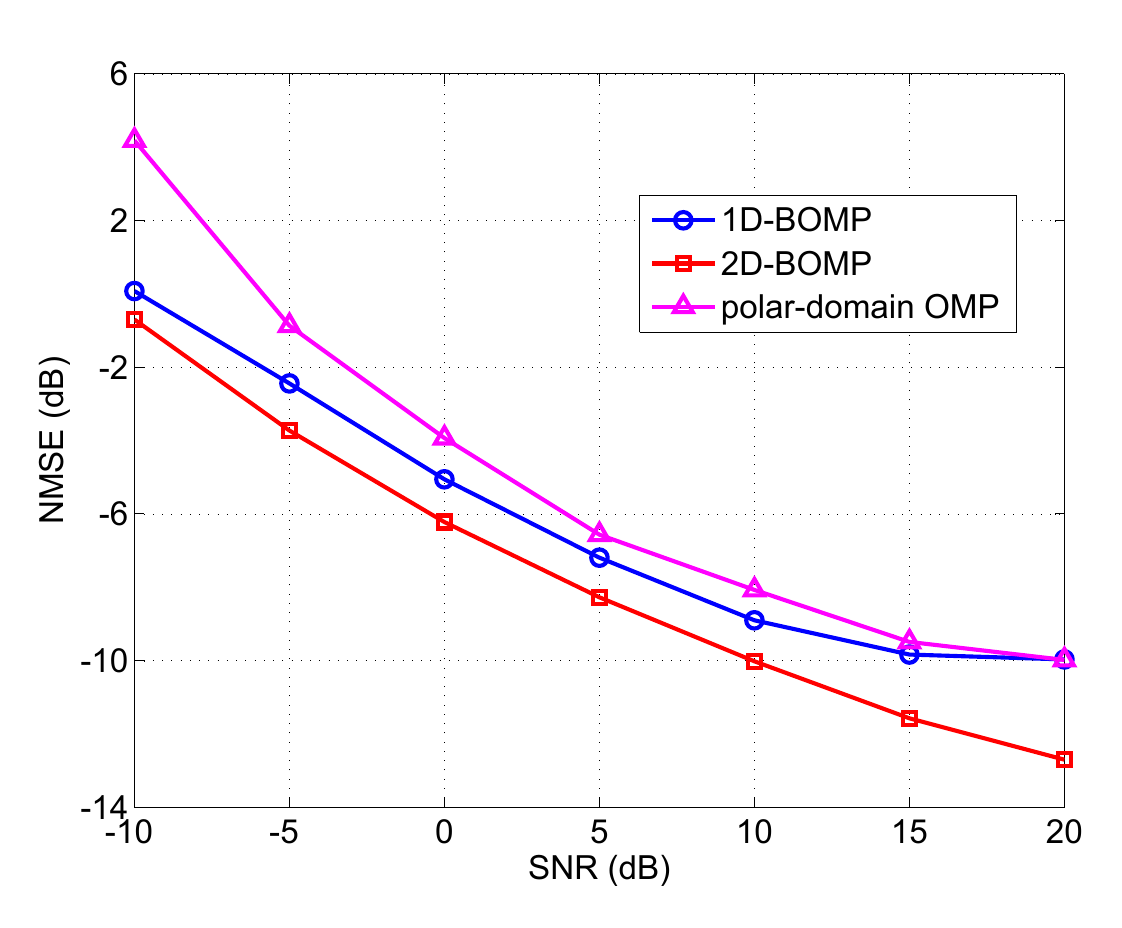}
  \caption{NMSEs of the comparative algorithms versus SNR when $T=1024$.}
  \label{fig1}
\end{figure}

\begin{figure}[htbp]
  \centering
  \includegraphics[width=0.35\textwidth]{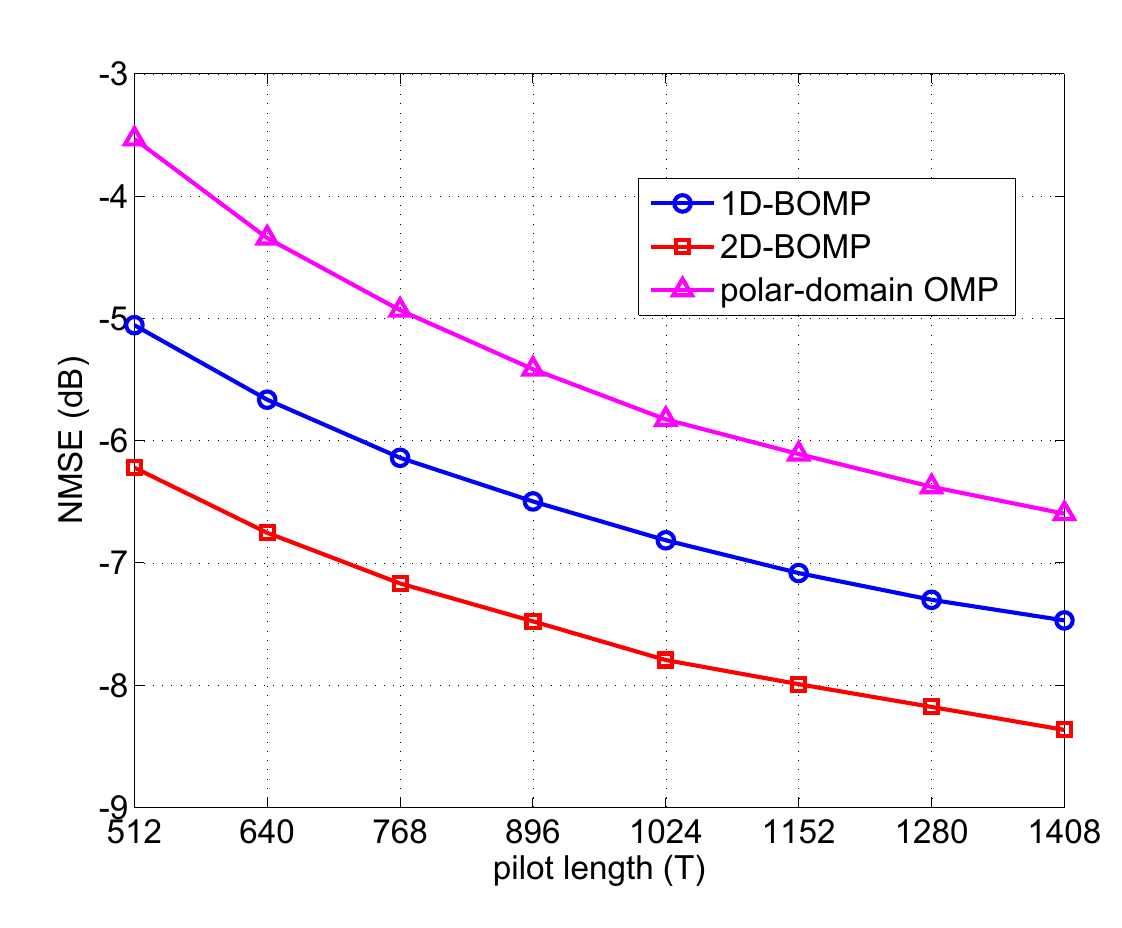}
  \caption{NMSEs of the comparative algorithms versus $T$ when $\text{SNR} = 5 \text{dB}$}
  \label{fig2}
\end{figure}

Fig.~\ref{fig1} illustrates the NMSE performance of the comparative algorithms as a function of the SNR, with a fixed pilot length of $T = 1024$. As the SNR increases from $-10$ dB to $20$ dB, the NMSE values for all schemes decrease monotonically, reflecting the expected reduction in estimation error as background noise diminishes. The proposed 2D-BOMP estimator consistently achieves the lowest NMSE across the entire SNR spectrum, outperforming both the 1D-BOMP baseline and the polar-domain OMP method. This performance improvement validates the effectiveness of the rectangular $2 \times 3$ coefficient grouping scheme used in 2D-BOMP, which captures localized channel structures more accurately than the consecutive groups of six coefficients implemented in conventional 1D-BOMP. 

Fig.~\ref{fig2} depicts the NMSE performance versus the pilot length $T$ (ranging from $512$ to $1408$) at a constant SNR of $5$ dB. As the pilot length increases, the estimation error for all algorithms drops steadily, demonstrating that supplying a greater number of pilot measurements enhances the system's ability to resolve channel coefficients. The 2D-BOMP algorithm maintains a consistent performance margin over 1D-BOMP and polar-domain OMP across all evaluated pilot lengths. Notably, 2D-BOMP achieves superior accuracy even at constrained pilot lengths, highlighting its high pilot efficiency and robustness for millimeter-wave and sub-THz massive MIMO channel estimation.

Overall, these simulation results confirm that structuring the coefficient selection in 2D-BOMP captures the underlying channel characteristics more accurately than standard 1D grouping or polar-domain OMP. This structural advantage translates to robust improvements in channel estimation precision across varying signal-to-noise ratios and pilot lengths.

\section{Conclusions}

This letter investigated near-field channel estimation for extremely large UPAs using 2D-BOMP via rectangular coefficient grouping. Our analysis showed that two-dimensional grouping reduces support uncertainty while preserving active dimensions under partition alignment. Numerical results demonstrated superior estimation accuracy over 1D grouping, though partition mismatch can diminish these performance gains. These insights highlight matched grouping as crucial for structured recovery and motivate adaptive partitioning and within-block refinement.

\bibliographystyle{IEEEtran}
\bibliography{refs}

\end{document}